\documentclass[%
 reprint,
nobibnotes,
 amsmath,amssymb,
 aps,
]{revtex4-1}

\usepackage{graphicx}
\usepackage{dcolumn}
\usepackage{bm}
\usepackage[titletoc,title]{appendix}
\usepackage{amsmath}

\usepackage[
top    = 1.8cm,
bottom = 1.5cm,
left   = 2cm,
right  = 2cm]{geometry}

\newcounter{mysubtable}

\begin{document}


\title{Consequences of minimizing pair correlations in fluids for dynamics, thermodynamics, and structure}

\author{Ryan B. Jadrich}
\affiliation{McKetta Department of Chemical Engineering, University of Texas at Austin, Austin, Texas 78712, USA}

\author{Beth A. Lindquist}
\affiliation{McKetta Department of Chemical Engineering, University of Texas at Austin, Austin, Texas 78712, USA}

\author{Jonathan A. Bollinger}
\affiliation{McKetta Department of Chemical Engineering, University of Texas at Austin, Austin, Texas 78712, USA}

\author{Thomas M. Truskett}
\email{truskett@che.utexas.edu}
\affiliation{McKetta Department of Chemical Engineering, University of Texas at Austin, Austin, Texas 78712, USA}

\date{\today}

\begin{abstract}

\noindent Liquid-state theory, computer simulation, and numerical optimization are used to investigate the extent to which positional correlations of a hard-sphere fluid--as characterized by the radial distribution function and the two-particle excess entropy--can be suppressed via the introduction of auxiliary pair interactions. The corresponding effects of such interactions on total excess entropy, density fluctuations, and single-particle dynamics are explored. Iso-\(g\) processes, whereby hard-sphere-fluid pair structure at a given density is preserved at higher densities via the introduction of a density-dependent, soft repulsive contribution to the pair potential, are considered. Such processes eventually terminate at a singular density, resulting in a state that--while incompressible and hyperuniform--remains unjammed and exhibits fluid-like dynamic properties. The extent to which static pair correlations can be suppressed to maximize pair disorder in a fluid with hard cores, determined via direct functional maximization of two-body excess entropy, is also considered. Systems approaching a state of maximized two-body entropy display a progressively growing bandwidth of suppressed density fluctuations, pointing to a relation between ``stealthiness'' and maximal pair disorder in materials.

\end{abstract}

\pacs{Valid PACS appear here}
\maketitle





 \section{Introduction}
\label{introduction}

Hard-sphere (HS) particles provide a basic structural model for a wide variety of particulate systems ranging from the macroscale (e.g., packings of ball bearings) to the microscale (e.g., colloidal suspensions or metallic glasses)~\cite{jamming_review,rcp_colloids,exp_colloid_hs_I,exp_colloid_hs_II}. At the microscale, thermal motion is the equilibrium configuration sampling mechanism, whereas randomly applied external perturbations (e.g., vibrations) can play an analogous role for macroscale systems. What is common across scales is the fundamental role of excluded volume due to the hard cores, which intrinsically limits the available configurational space of the system via strict geometric constraints. While such constraints have been extensively studied to gain insights into jamming phenomena~\cite{PhysRevLett.84.2064,PhysRevE.68.011306,jamming_review,  jamming_polytope, rcp_reproducible, universal_jamming} and the glass transition~\cite{exp_colloid_hs_I,exp_colloid_hs_II,hs_glass_theory, poly_hs_glass_theory,binary_hs_ultradense}, there is still much to be learned regarding the extent to which the hard-core interaction limits the possible structural arrangements of the particles--and hence the properties--of fluid systems. 

Perhaps the most important measure of structure for homogeneous fluids is the radial distribution function, \(g(r)\). The product $\rho g(r) d {\bf r}$ quantifies the expected number of particle centers to be found in a shell of differential width $dr$ a distance $r=|{\bf r}|$ away from a given particle center in a fluid of number density $\rho$. The core prevents particles from approaching closer than a particle diameter $d$; i.e., $g(r)=0$ for $r<d$. In dense fluids, packing constraints provided by the cores also give rise to a decaying oscillatory structure in $g(r)$ for $r>d$, signifying the coordination shells that naturally form around the particles. The coordination-shell structure quantified by the radial distribution function is critical for the thermodynamic properties (e.g., the equation of state) of the fluid, but it can influence transport properties as well~\cite{anomalous_coord_shells,tracer_scaling}.
   
One question of fundamental interest and potential practical importance for materials design~\cite{ST_inv_des_review,AJ_inv_des_review} is how much `pair disorder' can be accommodated by a fluid of  particles with hard cores of diameter $d$ at a given packing fraction \(\phi\)?\footnote[1]{The packing fraction of $D$-dimensional spherical particles of diameter $d$ is given by $\phi=v_D(d)\rho$, where $v_D(d)=\pi^{D/2}(d/2)^D/\Gamma(1+D/2)$.} In other words, to what extent is it possible to suppress the core-induced oscillations in $g(r)$, e.g., via incorporation of auxiliary pair interactions?
   
A particularly suitable metric for pair disorder in equilibrium systems, and the one adopted in this paper, is the two-body contribution that emerges in the multi-particle expansion of the excess (i.e., configurational, over ideal gas) entropy, $s_2$~\cite{s2_I,s2_II}.  For a fluid of $D$-dimensional particles with hard-core diameter $d$ and isotropic pair interactions, this quantity can be expressed in \emph{dimensionless} form, where Boltzmann's constant, $k_{\text B}$, is absorbed in the definition, as
\begin{equation}\label{eqn:pair_excess_entropy}
\begin{split}
   s_{2}=-2^{D-1}\phi \bigg[1+\dfrac{D}{d^{3}} &\int_{d}^{\infty} dr r^{D-1}\\
   &\times \{g(r)\text{ln}g(r)-g(r)+1\}\bigg]
\end{split}
\end{equation}\normalfont
In Equation~\ref{eqn:pair_excess_entropy}, the first term in the sum, $-2^{D-1}\phi$, is trivially due to the excluded volume of the hard core, while the second term accounts for any coordination-shell structure present in $g(r)$ for $r>d$.  In seeking to find systems that maximize $s_2$, it is important to ensure that known requirements for realizability are met. Here, we follow convention and require the static structure factor \(S(k)\equiv 1+\rho \hat{h}(k)\), related to $g(r)$ via the Fourier transform (FT) $\hat{h}(k) \equiv$ FT$[g(r)-1]$, to be positive definite for all \(k\).\footnote[2]{Positive definiteness of \(S(k)\) is a practical and strong, necessary~\cite{liquid_state_theory} condition for realizability of a given radial distribution function at given $\phi$.

More constraints are likely required to guarantee realizability of a packing, though understanding the nature and importance of these constraints remains an active area of research~\cite{sk_constraints}.} Thus, we formulate the two-body disorder maximization problem of interest as
\begin{equation}\label{eqn:positive_definite}
\begin{split}
& \underset{g(r)}{\text{max}}[s_{2}] \\
& g(r)=0, \ \ r\leq d \\
& S(k)\geq 0, \ \  0\leq k < \infty
\end{split}
\end{equation}\normalfont
For packing fractions $\phi \leq 2^{-D}$, the analytical solution of Equation~\ref{eqn:positive_definite} is simply the unit Heaviside step function \(g(r)=H(r-d)\), i.e., identical to the limiting radial distribution of the HS fluid as $\phi \rightarrow 0$, with $s_2=-2^{D-1}\phi$. In other words, a radial distribution function without \emph{any} oscillatory coordination-shell structure appears realizable for fluids of particles with hard cores under these conditions. Precisely at \(\phi=2^{-D}\), the solution \(g(r)=H(r-d)\) leads to hyperuniform structure (long ranged density fluctuations are suppressed, \(S(0)=0\)~\cite{isog_and_hyperuniform,hyperuniform_I}) and thus the system is incompressible\footnote[3]{Hyperuniformity is characteristic of close-packed crystalline states, though disordered isotropic materials (i.e., maximally randomly jammed packings and the systems studied in this work) can also meet this criterion.~\cite{isog_and_hyperuniform,hyperuniform_I}}. For packing fractions \(\phi > 2^{-D}\), the step function solution would violate the realizability constraint, i.e., exhibiting \(S(k)<0\) for some range of \(k\)~\cite{isog_I,isog_II,isog_and_hyperuniform}. Hence, the maximally disordered pair structure for \(\phi > 2^{-D}\) must generally be obtained via a numerical solution of Equation~\ref{eqn:positive_definite}. 

The low-density, analytical solution to the $s_2$ maximization problem of Equation~\ref{eqn:positive_definite} described above also provides information pertaining to a so-called iso-\(g\) [i.e., iso-\(g^{(2)}(r)\)] process~\cite{isog_I,isog_II,isog_and_hyperuniform}.
An iso-\(g\) process is one where an equilibrium system at an initial packing fraction, \(\phi_{0}\) (in the above example, the HS fluid at $\phi_0=0$), can realize an identical radial distribution function at a higher packing fraction $\phi$ by adding a $\phi$-dependent auxiliary pairwise interaction which counteracts any pairwise structural evolution that would have otherwise occurred. For a given iso-\(g\) process, there is also a singular packing fraction, $\phi_s$ (in the above example, $\phi_s=2^{-D}$), above which the pair correlation of the original state can no longer be realized because it would unphysically result in $S(k) < 0$ for some values of $k$.




Irrespective of the chosen value for \(\phi_{0}\), iso-\(g\) processes initiating from an equilibrium HS structure (1) necessarily grow in auxiliary, soft pair interactions external to the hard core at $\phi>\phi_0$ to prevent pair structure evolution and (2) terminate at an upper singular packing fraction \(\phi_{\text{s}}\) at which \(S(0)= 0\), corresponding to an incompressible, hyperuniform state.\footnote[4]{The singular density is a direct consequence of halting pair structure evolution and can be understood as follows. Holding \(g(r)\) constant is equivalent to fixing \(\hat{h}(k)\), a quantity that exhibits a global negative minimum at \(k=0\) for hard spheres (typical of any purely repulsive system). Thus, there always exists a singular density, \(\rho_{\text{s}}\), at which \(S(0)\equiv 1+\rho_{\text{s}} \hat{h}(0)=0\).} Since the pair structure of such an iso-$g$ process is, by definition, identical that of the HS fluid at the lower initial packing fraction ($\phi_0 < \phi_s$), its pair disorder (e.g., as measured by $s_2$) at $\phi_s$ is always higher than that of an equi-density equilibrium HS fluid, but less than or equal to that obtained by a formal maximization of $s_2$ via Equation~\ref{eqn:positive_definite}.

Since an iso-\(g\) process is technically simpler to analyze than a full \(s_{2}\) maximization, the former is a helpful construct for studying how the enhancement of two-body disorder of a hard-core fluid affects the total \emph{dimensionless} excess entropy, \(s\) (inclusive of all many-body correlation contributions) and other related properties. Na\"{i}vely, one might anticipate that muting pair structure would also increase the overall structural disorder of the system, which would conceptually align with factorization approximations in equilibrium statistical mechanics including the Kirkwood superposition approximation that estimates the three-body distribution according to:  \(g^{(3)}(\boldsymbol{r}_{1},\boldsymbol{r}_{2},\boldsymbol{r}_{3})\approx g(\boldsymbol{r}_{1}-\boldsymbol{r}_{2})g(\boldsymbol{r}_{1}-\boldsymbol{r}_{3})g(\boldsymbol{r}_{2}-\boldsymbol{r}_{3})\)~\cite{liquid_state_theory,mcquarrie}. However, upon further examination, this logic directly contradicts expectations based on thermodynamics. For example, the auxiliary interactions that are added to the hard-core potential to preserve structure in an iso-$g$ process, as discussed more formally in Section~\ref{results:step_isog}, must necessarily cause a \emph{decrease} in \(s\) compared to the HS reference state (despite the relative increase in $s_2$). The magnitude of this decrease, which is due to corresponding (and more than compensating) enhanced structure in higher-body static correlations compared to the HS reference fluid, has not yet been explored and is one focal point of this work. 

Probing the corresponding changes of \(s\) and \(s_{2}\) along iso-\(g\) processes is also of interest given the empirically observed links between these quantities and dynamic properties of fluids~\cite{rosenfeld_I,rosenfeld_II,dzugutov_I,doi:10.1021/jp071369e}. 
For many fluid systems, changes that increase excess entropy (or its two-body contribution) result in faster dynamical relaxation processes and vice-versa, even for systems that exhibit ``anomalous'' trends in dynamics~\cite{1.2409932,anomalous_gauss,anomalous_coord_shells,PhysRevE.80.061205,1.3256235,1.3559676} with respect to quantities like number density. 
While \(s\) and \(s_{2}\) are typically strongly correlated (the latter is usually the dominant contribution to the former), they would yield conflicting dynamical predictions as a function of $\phi$ for iso-\(g\) processes starting from a HS state at $\phi_0$. 
From the perspective of \(s_{2}\), an iso-\(g\) process should produce enhanced dynamics relative to the HS fluid at the same packing fraction. This at odds with expectations based on behavior of \(s\), a quantity which is \emph{necessarily} lowered relative to the hard sphere-fluid via an iso-\(g\) process. While the magnitude of the \(s\) reduction along an iso-\(g\) path is unknown, a large decrease (and implied dynamical slowdown) might seem a reasonable expectation given the tendency toward incompressibility of an iso-\(g\) fluid upon approach to the corresponding singular packing fraction $\phi_s$--a topic we address in this paper. Along similar lines, strong correlations between dynamical relaxation times and the excess compressibility factor have been observed in simulations of fluids with hard cores (see, e.g., \cite{1.1288804,jp902934x}). Whether such correlations also hold for iso-$g$ processes, where they would predict a pronounced dynamic slowdown approaching $\phi_s$, remains an open question and a stringent test of their generality.

Also critical to understanding the static and dynamic properties displayed by systems of an iso-\(g\) process are the $\phi$- (and $\phi_0$-) dependent auxiliary interactions which must be incorporated to keep the pair structure constant as packing fraction is increased. These interactions are known to be of a repulsive Yukawa form at large inter-particle separations and strongly believed, in accordance with predictions from approximate integral equation theories, to be of a simple repulsive ramp-like form near contact for the step-function HS iso-\(g\) process~\cite{isog_I,isog_II,isog_and_hyperuniform}. Interestingly, a ramp-like interparticle potential was also recently shown to remove the coordination-shell structure of a tracer solute in a HS solvent~\cite{tracer_scaling,1.4916053}, and a ramp-like fluid-wall interaction is known to similarly `flatten' the density profile of a confined HS fluid~\cite{PhysRevLett.100.106001}. To clarify the above issues, we seek to address the general accuracy of integral equation theory predictions for an iso-\(g\) process, as tested by molecular dynamics (MD) simulations, and confirm the possibly generic ramp-like interparticle potential form that apparently emerges for iso-\(g\) processes that initiate from HS fluid reference states. 

Moreover, while iso-\(g\) processes provide a simple means by which to find interactions that increase \(s_{2}\) (relative to the HS fluid at the same packing fraction) and hence help to explore the ensuing thermodynamic, dynamic, and potential interaction consequences, we also want to examine the impact of formally maximizing two-body disorder at arbitrary density (see Equation~\ref{eqn:positive_definite}). Here, we tackle this problem via numerical optimizations performed with a genetic algorithm to reveal the absolute limits of two-body disorder in hard-core fluids. Of key interest is the residual structure that remains after $s_2$ maximization and the potentially unique changes in properties that the corresponding fluids display. In particular, we demonstrate a unique correspondence between minimizing pair disorder and suppressing structure not only \(k=0\) in \(S(k)\), but a continuous range in \(k\) wavevectors. Systems with a range of complete suppression in \(S(k)\) are termed ``stealthy'' since they are transparent (do not scatter) over a continuum of \(k\)~\footnote[5]{``Stealthy'' packings can be both ordered and disordered structurally. For the former, perfect crystals (no lattice displacements due to thermal motion or otherwise) are ``stealthy'' at all wavevectors not associated with Bragg scattering. For the latter, special disordered \emph{point} patterns have been constructed that specifically suppress a span of low \(k\) structure ~\cite{stealthy_hyperuniform_I,stealthy_hyperuniform_II}.}.

The balance of this paper is organized into two primary Sections. Section~\ref{methods} describes our liquid-state theoretical calculations, MD simulations, and numerical optimizations. Section~\ref{results} outlines our thermodynamic, dynamic and auxiliary potential predictions for an array of iso-\(g\) processes while also discussing our \(s_{2}\) maximization results. Section~\ref{conclusions} concludes the paper and discusses some interesting future directions. We also provide a review in Appendix A of iso-\(g\) processes and the corresponding mathematical framework for their analysis (used implicitly and referenced throughout the text).

\section{Methods}
\label{methods}

\subsection{Integral equation theory}
\label{methods:IET}

Integral equation theory (IET) is a powerful tool of equilibrium statistical mechanics that is most frequently used for the forward prediction of \(g(r)\) and related thermodynamic properties based on knowledge of the dimensionless pair potential between particles, \(u(r)\)~\cite{liquid_state_theory,mcquarrie} [implicitly contains the inverse thermal energy $\beta = (k_{\text B} T)^{-1}$ where $T$ is temperature and $k_{\text B}$ is Boltzmann's constant]. However, IET can also be used for the inverse problem--as studied for three-dimensional fluid systems in this paper--of deriving the necessary \( u(r)\) to form a system exhibiting a desired target \(g(r)\). IET employs a formally exact partitioning of the total correlation function, \(h(r)\equiv g(r)-1\), into direct and multibody contributions via the direct correlation function, \(c(r)\), and the Ornstein-Zernike (OZ) relation
\begin{equation}\label{eqn:OZ}
   h(r)=c(r)+\rho \int_{}^{} c(r')h(|\mathbf{r}-\mathbf{r}'|) d\mathbf{r}
\end{equation}\normalfont

The OZ relation can be used together with approximate closures relating \( u(r)\), \(g(r)\) and \(c(r)\). Three popular closures are explored in this paper, namely, the hypernetted chain (HNC)~\cite{liquid_state_theory,mcquarrie,percus_PY_HNC}, Percus-Yevick (PY)~\cite{liquid_state_theory,mcquarrie,percus_PY_HNC}, and Martinov-Sarkisov (MS)~\cite{MS_closure} approximations
\begin{equation}\label{eqn:HNC_closure}
    u(r)\equiv g(r)-1-\text{ln}[g(r)]-c(r),\ \ \ \text{HNC}
\end{equation}\normalfont
\begin{equation}\label{eqn:PY_closure}
    u(r)\equiv \text{ln}[1-c(r)/g(r)],\ \ \ \text{PY}
\end{equation}\normalfont
\begin{equation}\label{eqn:MS_closure}
    u(r)\equiv \sqrt{1+2[g(r)-c(r)-1]}-\text{ln}[g(r)]-1,\ \ \ \text{MS}
\end{equation}\normalfont
We also consider the random phase approximation (RPA)~\cite{liquid_state_theory,mcquarrie}
\begin{equation}\label{eqn:RPA_closure}
    u(r)\equiv c_{0}(r)-c(r),\ \ \ \text{RPA}
\end{equation}\normalfont
where \(c_{0}(r)\) is an assumed known (here, hard-sphere) direct correlation function. To determine \(c_{0}(r)\), we use the HS structure predicted from a forward IET calculation with the modified Verlet (MV) closure~\cite{VM_closure}
\begin{equation}\label{eqn:MV_closure}
\begin{split}
   g_{0}(r)= \text{exp}  
   & \bigg[h_{0}(r)-c_{0}(r) \\ 
   & -\dfrac{[h_{0}(r)-c_{0}(r)]^{2}}{2+(8/5)[h_{0}(r)-c_{0}(r)]} \bigg],\ \ \ \text{MV}
\end{split}
\end{equation}\normalfont
Combining Equation~\ref{eqn:OZ} with one of Equations~\ref{eqn:HNC_closure}-\ref{eqn:RPA_closure}, either \(g(r)\) or \( u(r)\) can be obtained given knowledge of the other.

For IET, the inverse problem is technically simpler than the forward problem as numerical solution is not necessary for most simple closures like Equations~\ref{eqn:HNC_closure}-\ref{eqn:MS_closure}. Instead, solution follows by inputting $\hat{h}(k)$, a FT of the known \(h(r)\), into Eqn.~\ref{eqn:OZ} and then solving for \(\hat{\gamma}(k)\equiv \hat{h}(k)-\hat{c}(k)\)
\begin{equation}\label{eqn:OZ_FT}
   \hat{\gamma}(k)=\dfrac{\rho \hat{h}^{2}(k)}{1+\rho \hat{h}(k)}
\end{equation}\normalfont
which, after an inverse Fourier transform, yields the direct correlation function via \(c(r)=h(r)-\gamma(r)\)~\footnote[6]{One solves for \(\hat{\gamma}(k)\), rather than \(\hat{c}(k)\), in order to avoid numerical error induced by ringing about the core discontinuity upon transforming back to \(r\)-space}. Provided the \(c(r)\) corresponding to the target \(g(r)\), Eqns.~\ref{eqn:HNC_closure}-\ref{eqn:RPA_closure} can be used to determine the corresponding \( u(r)\).

Additionally, within the HNC and RPA approximations, the \emph{dimensionless} excess Helmholtz free energy per particle \(f\) can be calculated using~\cite{liquid_state_theory}
\begin{equation}\label{eqn:HNC_free_energy}
\begin{split}
&f=-2\pi \rho \int_{0}^{\infty}dr r^{2}\Big[c(r)-h(r)^{2}/2\Big] \\
&+\dfrac{1}{4\pi^{2}\rho}\int_{0}^{\infty}dkk^{2}\Big[\rho \hat{c}(k)+\text{ln}[1-\rho \hat{c}(k)]\Big], \ \ \text{HNC}
\end{split}
\end{equation}\normalfont
and
\begin{equation}\label{eqn:RPA_free_energy}
\begin{split}
f=f_{0}
&+2\pi \rho \int_{0}^{\infty}dr r^{2}g_{0}(r) u(r) \\
&-\dfrac{1}{4\pi^{2}\rho}
\int_{0}^{\infty}dkk^{2}\Big[\rho S_{0}(k) \hat{u}(k) \\
&+\text{ln}[1-\rho S_{0}(k) \hat{u}(k)]\Big], \ \ \text{RPA}
\end{split}
\end{equation}\normalfont
respectively, where \(f_{0}\) is the HS reference, dimensionless excess Helmholtz free energy per particle (we employ the accurate Carnahan-Starling form~\cite{liquid_state_theory,carnahan_starling}). Free energy calculations provide \(s\) via \(s=e-f\) where \(e\equiv (\rho/2)\int_{}^{}d\textbf{r}g(r) u(r)\) is the dimensionless excess (potential) internal energy per particle.

\subsection{Molecular dynamics simulations}
\label{methods:MD}

For selected states, we conduct MD simulations of monodisperse particles governed by pair potentials \( u(r) =  u_{\text{WCA}}(r) +  u_{\text{F}}(r)\). Here, \( u_{\text{WCA}}(r)\) is a normalized, steeply repulsive Weeks-Chandler-Andersen (WCA) interaction representing a hard-core exclusion volume~\cite{liquid_state_theory,ChandlerWeeksAndersenScience1983}, which is given by \( u_{\text{WCA}}(r) = 4([d/r]^{48} - [d/r]^{24}) + 1\) for \(r\leq2^{1/24}d\) and \( u_{\text{WCA}}(r)=0\) for \(r>2^{1/24}d\). The remaining term, \( u_{\text{F}}(r)\), represents the density-dependent ``flattening'' potentials derived via IET to actualize iso-\(g\) processes, where this term is set to \( u_{\text{F}}(r)=0\) when approximating HS fluids.~\footnote[7]{The full pair potentials \( u(r)\) obtained in the inverse IET framework are defined by literal HS potentials (\( u_{\text{HS}} = \infty\) for \(r < d\) and \( u_{\text{HS}} = 0\) otherwise) combined with \( u_{\text{F,IET}}(r)\) potentials that are non-zero only for \(r \geq d\). To approximate \( u(r)\) with continuous potentials, we superimpose \( u_{\text{WCA}}(r)\) and \( u_{\text{F}}(r)\), where \( u_{\text{F}}(r)\) is the potential \( u_{\text{F,IET}}(r)\) extended linearly for \(r\leq d\) according to the derivative \(\text{d}[ u_{\text{F,IET}}(r)]/\text{d}r\) at \(r\rightarrow d^{+}\). This ensures smoothness of the superimposed potential at all \(r\).} When simulating systems with non-zero \( u_{\text{F}}(r)\), we choose the cutoff distance \(r_{\text{c}}\) for \( u(r)\) such that the force at this distance \(F(r_{\text{c}}) = -\text{d}( u(r_{\text{c}}))/\text{d}r = 0.001\).

Using these potentials, we perform three-dimensional simulations in the canonical ensemble with periodic boundary conditions using GROMACS 4.6.5~\cite{HessJCTC2008}. We use an integration time-step of \(dt=0.001\sqrt{\beta d^{2}m}\) and fix temperature via a Nos\'{e}-Hoover thermostat with time-constant \(\tau=1000dt\). For HS systems, we simulate \(N=3000\) particles; for systems governed by non-zero \( u_{\text{F}}(r)\), we simulate between \(3000 \leq N \leq 6000\) particles, where \(N\) is proportional to the potential cutoff distance \(r_{\text{c}}\), which for our state points of interest ranges between \(3 \lesssim r_{\text{c}}/d \lesssim 13\).

To characterize pair correlations, we calculate the radial distribution function \(g(r)\) in the usual way from particle configurations; subsequently, we calculate \(S(k)\) using the identity \(\hat{h}(k)\equiv \text{FT}[g(r)-1]\). To characterize dynamics, we calculate the self-intermediate scattering function \(F_{\text{s}}(\textbf{k},t) = (1/N) \langle \sum_{j=1}^{N} \exp{\{-i\textbf{k}\cdot[\textbf{r}_{j}(t) - \textbf{r}_{j}(0)]\}}\rangle\), where \(\textbf{r}_{j}(t)\) is the instantaneous position of particle \(j\) at some lag-time \(t \geq 0\) relative to its position at \(t = 0\) and \(\langle ... \rangle\) represents an ensemble average. Given the systems are isotropic, we average statistics across all three spatial directions to obtain \(F_{\text{s}}(k,t)\). When calculating this quantity, we take care to set up simulations with box lengths \(L\) appropriate for obtaining \(F_{\text{s}}(k=2\pi/d,t)\), where examining this wavevector magnitude (corresponding to a real-space distance of \(d\)) characterizes the dynamic decay of correlations due to single-particle diffusion.

\subsection{Genetic algorithm maximization of $s_2$}
\label{methods:GA}

As described in the Introduction, Equation~\ref{eqn:positive_definite} has an analytical solution for $\phi \le 1/8$ in $D=3$. The radial distribution function with $S(k) \geq 0$ for all $k$ that maximizes $s_2$ under these conditions is the unit step function $g(r)=H(r-d)$.  For $\phi > 1/8$, we used a genetic algorithm~\cite{genetic_algorithms_I,genetic_algorithms_II} to numerically solve Equation~\ref{eqn:positive_definite}. In order to avoid restriction to any particular functional form, we directly optimized the radial distribution function at 50 different values of $r$, interpolating between the points via a cubic spline to construct the full correlation function. Based on preliminary optimizations, the knots for the spline were spaced unevenly (weighted towards smaller values of $r$), giving more flexibility to the \(g(r)\) near contact. To enforce the hard-core constraint, \(g(r)\) for \(r < d\) was fixed to be zero. Additionally, $g(r)$ for \(r \geq 6d\) was set to unity. Between these two bounds, \(g(r)\) was optimized, and the associated \(S(k)\) was calculated using the FFTW program~\cite{fftw}. If \(S(k)\) was not positive definite, then the minimum value of \(S(k)\), weighted by a positive empirical factor ($\alpha$), was added as a penalty to the figure of merit, 
and optimization cycles proceeded with an increasing \(\alpha\) until the constraint was satisfied.

The optimization was performed with a micro-genetic algorithm with a population size of 5 using tournament selection, 50\% uniform crossover and 50\% single point crossover, a mutation rate of 5\%, elitism, binary encoding, and each cycle comprised 50,000 steps.~\cite{genetic_algorithms_I,genetic_algorithms_II} In order to restrict parameter space such that a solution could be found efficiently, the randomly generated \(g(r)\) was smoothed once with a single triangular smoothing function. As a practical consideration, a genetic algorithm requires discretization of the solution space. As a consequence, several solutions exist that are very close in \(s_{2}\). Therefore, 10 independent optimizations were carried out at each density and level of discretization. Optimizations were performed iteratively, where the allowed precision of the \(g(r)\) values was incrementally increased as the range of values allowed in the optimization were concurrently decreased as determined from the spread of the data in the previous step. If this range was exceeded in a subsequent optimization, the range was widened in the previous step until all optimizations fell inside the predetermined range. Once a discretization in the \(g(r)\) of \(1.56\times10^{-4}\) was achieved, the 10 solutions were averaged to arrive at the final optimized \(g(r)\).

\section{Results}
\label{results}

\subsection{Preserving ideal structure via an iso-\(g\) process}
\label{results:step_isog}

\begin{figure}
  \includegraphics{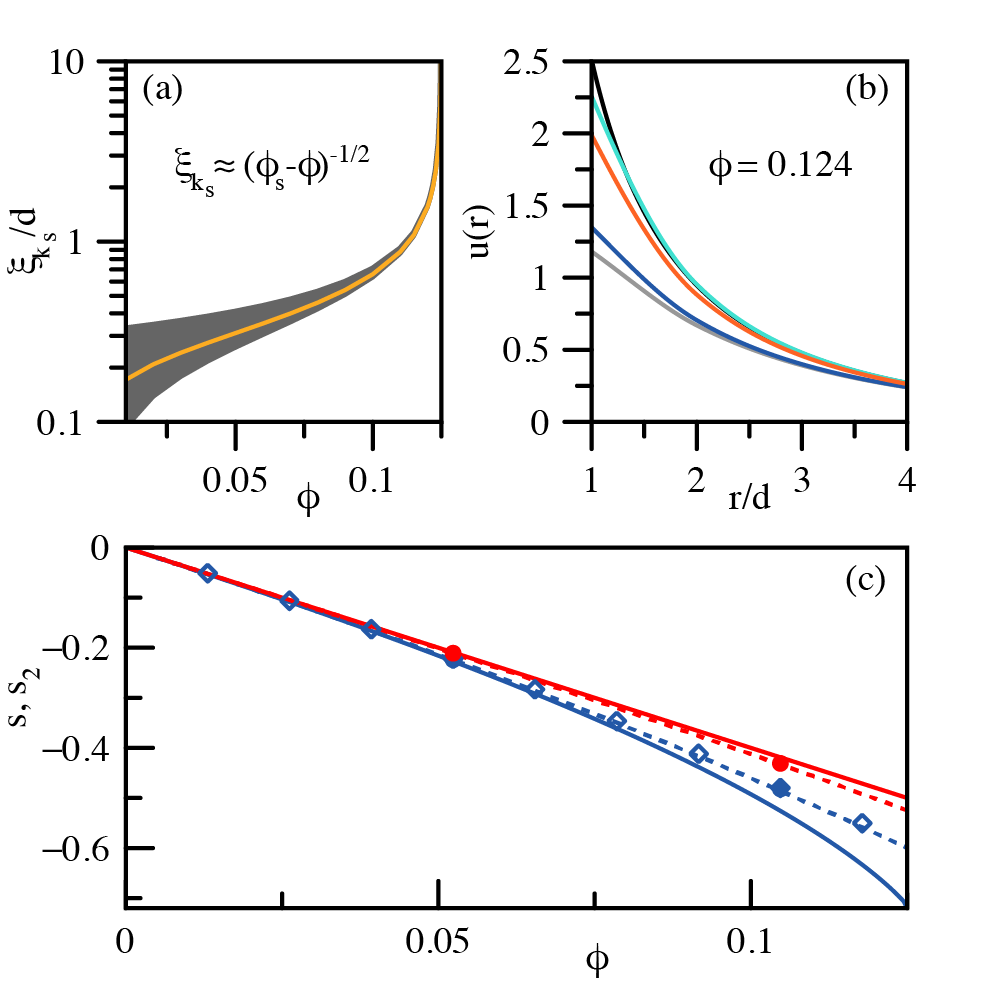}
  \caption{(a) The range of the auxiliary soft repulsion, $\xi_{k_{\text{s}}}$, required for the iso-\(g\) process initiated from the HS fluid at $\phi_0=0$, as function of packing fraction $\phi$. The solid yellow line represents the exact result from Eqn.~\ref{eqn:exact_step_cr} and the shaded region denotes the upper and lower bound approximations derived in Appendix A. (b) Representative soft auxiliary potentials near the singular packing fraction $\phi_s=1/8$ predicted using IET and (from bottom to top) PY, MS, RPA, and HNC closures as well as the asymptotic Yukawa form [derived in Appendix A]. (c) HNC predictions for full (lower blue curves) and two-body (upper red curves) excess entropy for this iso-g process [solid lines] as compared to the unperturbed hard sphere fluid [dashed lines]. Circles and diamonds are simulation results from ~\cite{s_s2_simulations} and ~\cite{s_simulations} respectively.}
  \label{sch:auxiliary_range}
\end{figure}

We begin our discussion by considering three-dimensional fluids at relatively low packing fractions (\(0 \leq \phi \leq 1/8\)), where--as mentioned in the Introduction--the radial distribution function that maximizes $s_2$ subject to a hard-core constraint corresponds to that of the HS fluid at vanishing density [i.e., a step function, $g(r)=H(r-d)$]. The range of equilibrium fluid states that can exhibit this ideal pair structure via the addition of auxiliary pair interactions to the hard-core potential
corresponds to an iso-\(g\) process (initiating from a HS fluid with density \(\phi_{0}=0\)) that terminates at the singular density \(\phi_{\text{s}}=1/8\), at which \(S(0)=0\). We examine this relatively simple case to address the following questions, with a focus on behavior at \(\phi_{\text{s}}\), where the most pronounced HS structure is suppressed. What is the nature of the auxiliary interactions required to eliminate the hard-core pair correlations for $r >d$? How are thermodynamic quantities, including the two-body and total excess entropy, affected by these changes? Are incompressible states that must exist at \(\phi_{\text{s}}\) dynamically slow or jammed in any sense?

%
%
As shown in Fig. 1, preventing pair correlation evolution as the the singular density is approached requires pronounced growth in the lengthscale of the auxiliary potential, which modifies system thermodynamics relative to that of bare hard spheres. First, as outlined in Appendix A, the large-$r$ asymptotic portion of the auxiliary potential [\( u(r)\propto \text{exp}(-r/\xi_{k_{\text{s}}})/r\)] becomes long ranged [\(\xi_{k_{\text{s}}}\rightarrow \infty\)] in a power law fashion [\(\xi_{k_{\text{s}}}\propto(\phi_{\text{s}}-\phi)^{-1/2}\)], serving as a sort of ``inverted'' critical phenomenon~\cite{isog_and_hyperuniform}. Here, the potential range can be calculated exactly from the imaginary portion, \(\xi_{k_{\text{s}}}=1/k_{\text{I}}\), of the \(k=0\) complex root in the denominator of \(\hat{c}(k)\) for the unit step \(g(r)\)
\begin{equation}\label{eqn:exact_step_cr}
    \hat{c}(k)=\dfrac{1}{\rho+\dfrac{k^{3}}{4k\pi d\text{cos}(kd)-4\pi \text{sin}(kd)}}
\end{equation}\normalfont
for which the solution (yellow line) is shown in Fig.~\ref{sch:auxiliary_range}a. Also shown in Fig.~\ref{sch:auxiliary_range}a is a shaded region denoting the space between two approximations derived in Appendix A (Equations A3 and A4). These generally applicable analytical forms (1) serve as useful upper and lower bounds and (2) indicate the onset of ``inverted'' critical-like behavior by their collapse. Collapse is realized once the potential range grows larger than the particle diameter, an intuitive result given that \(d\) is the only relevant length scale to exceed. Some IET auxiliary potential predictions within this critical limit are presented in Fig.~\ref{sch:auxiliary_range}b along with the asymptotic contribution (Equation~\ref{eqn:yukawa_k0}). While all four theories correctly merge to the asymptotic result as \(r \rightarrow \infty\) (not shown) and display a ramp-like form near contact, notable quantitative differences are apparent. The accuracy of each theory is tested later in this Section, but universal amongst each is the predicted divergent excess internal energy, \(e\equiv (\rho/2)\int_{}^{}d\textbf{r}g(r) u(r)\), that arises upon approach to \(\phi_{\text{s}}\). More precisely, \(\lim_{\xi_{k_{\text{s}}\to\infty}}e\propto \rho \xi_{k_{\text{s}}}^{2}\). The Helmholtz energy \(f\) also diverges upon approach to \(\phi_{\text{s}}\) but, as we show below, \(s=e-f\) remains finite implying an exact cancellation. Ultimately, it is the divergence in \(e\), and not entropic factors contained within \(f\), that drives the vanishing of the compressibility at \(\phi_{\text{s}}\).

Along an iso-\(g\) process initiating from a HS fluid state at $\phi_0$, the `pair disorder' [characterized by \(s_{2}\)] progressively increases with $\phi$ relative to that of the HS fluid at the same packing fraction. But, what are the consequences for the overall configurational disorder [quantified via \(s\)]? This can be understood by beginning with the rigorous Gibbs-Bogoliubov (GB) inequality~\cite{liquid_state_theory}
\begin{equation}\label{eqn:GB_inequality_I}
    f_{0}+\dfrac{\rho}{2}\int{}{}d\textbf{r}g(r) u(r) \leq f \leq f_{0}+\dfrac{\rho}{2}\int{}{}d\textbf{r}g_{0}(r) u(r)
\end{equation}\normalfont
where \(f_{0}\) is the dimensionless, excess Helmholtz free energy per particle of the HS fluid, \(g(r)\) is the radial distribution function of the iso-\(g\) process with $\rho$-dependent auxiliary potential \(u(r)\), and \(g_{0}(r)\) is the radial distribution function of the HS fluid. Using the relations, \(f=e-s\), \(e\equiv (\rho/2)\int_{}^{}d\textbf{r}g(r) u(r)\) and \(f_{0}=-s_{0}\) in Equation~\ref{eqn:GB_inequality_I} yields
\begin{equation}\label{eqn:GB_inequality_II}
    s_{0}-\dfrac{\rho}{2}\int{}{}d\textbf{r}[g_{0}(r)-g(r)] u(r) \leq s \leq s_{0}
\end{equation}\normalfont
proving that $s$ necessarily decreases along an iso-\(g\) process (relative to the equidensity HS fluid). This result also makes intuitive sense, given that one applies auxiliary interactions in the iso-$g$ process to modify the structure of a fluid whose equilibrium state is, by definition, one of maximum entropy subject to the hard-core excluded volume constraint. 

Importantly, the decrease in \(s\) is \emph{finite bounded} because \(g_{0}(r)-g(r)\) will generally decay exponentially with $r$--preventing unbounded growth of the integral in Equation~\ref{eqn:GB_inequality_II}, despite the growing range of \( u(r)\). More specific to iso-\(g\) processes initiating from a HS fluid state, we expect the decay in the integral to be quite strong since HS radial distribution functions decay rapidly with $r$ at moderate densities~\cite{liquid_state_theory}, suggesting a fairly tight lower bound on \(s\). 


Two consequences for iso-\(g\) processes follow from Equation~\ref{eqn:GB_inequality_II}: (1) approach to the incompressible fluid at $\phi_s$ is qualitatively different from approach to an incompressible, jammed or close-packed state, which would exhibit a negative divergence in \(s\)~\cite{glasses_s2}~\footnote[8]{This reflects the squeezing out of vibrational motion (entropy) in locally trapped configurations or ``basins of attraction''. It is in these configurational basins that a hard sphere system would become dynamically trapped upon compression~\cite{jamming_review,hs_glass_theory} (neglecting so called ``rattler'' particles).}, and (2) fluid relaxation times might be predicted to increase \emph{or} decrease depending on whether \(s\) or \(s_{2}\) correlates with dynamics. To assess the expected changes in \(s_{2}\) and \(s\) along the unit-step function iso-\(g\) process from \(\phi_{0}=0 \rightarrow \phi_{\text{s}}=1/8\), we calculate these properties based on \( u(r)\) and \(g(r)\) obtained from HNC frameworks. As shown in Fig.~\ref{sch:auxiliary_range}c, a nominal increase in \(s_{2}\) over hard spheres is seen. For \(s\), the change is opposite in magnitude and slightly larger than that observed for \(s_{2}\)--though in an absolute sense the changes are still small, even at the singular point. Given the modest changes in $s$ and $s_2$ along the iso-$g$ process shown in Fig.~\ref{sch:auxiliary_range}, one would not anticipate a pronounced change in dynamics. However, such predictions contradict expectations based on dynamical correlations with the excess compressibility factor \cite{1.1288804,jp902934x}), which would anticipate a pronounced dynamical slowdown approaching $\phi_s$.



\begin{figure}
  \includegraphics
  {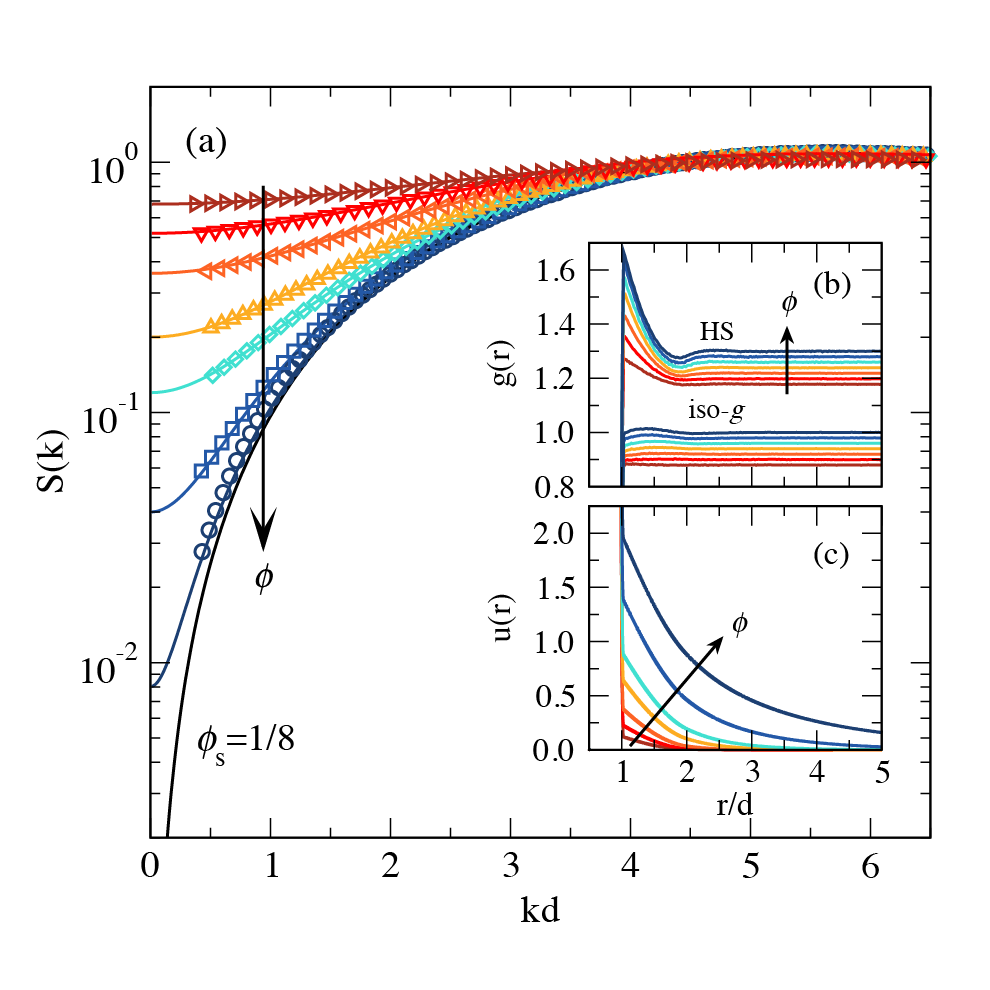}
  \caption{(a) Structure factors \(S(k)\) for the iso-\(g\) process spanning \(\phi_{0}=0 \rightarrow \phi_{\text{s}}=1/8\), including intermediate packing fractions \(\phi = 0.040,0.060,0.080,0.100,0.110,0.120,\) and \(0.124\). Lines correspond to target step-function \(g(r)\) profiles matching HS structure at \(\phi_{0}=0\) and symbols correspond to simulation results using potentials \( u(r)\) derived at each \(\phi\) to suppress oscillatory pair structure. (b) Radial distribution functions \(g(r)\) from simulations of HS systems and iso-\(g\) systems, vertically offset for clarity. (c) Full interaction potentials \( u(r) =  u_{\text{WCA}}(r) +  u_{\text{F}}(r)\) used to realize iso-\(g\) systems in panels (a) and (b).}
  \label{sch:Skgrur-lowdens}
\end{figure}

Given these analytical results, we next demonstrate via MD simulations that interactions such as those shown in Fig.~\ref{sch:auxiliary_range}b can indeed minimize pair correlations and that upon approaching the singular density, the fluid states approach incompressibility while still exhibiting finite relaxation times comparable to equidensity HS. To wit, Fig.~\ref{sch:Skgrur-lowdens}a shows structure factors \(S(k)\) for simulated iso-\(g\) state points (no HS results shown) between \(0 < \phi < 1/8\), while Fig.~\ref{sch:Skgrur-lowdens}b shows corresponding RDFs for both iso-\(g\) and HS simulations. Here, the iso-\(g\) systems are governed by the potentials \( u(r)\) shown in Fig.~\ref{sch:Skgrur-lowdens}c, which are a 3:1 weighted combinations of potentials derived via HNC and PY closures for target step-function \(g(r)\) profiles.~\footnote[9]{This mixture ratio is based on preliminary simulations, where we observe that potentials from the HNC and PY closures over- and under-suppress structural correlations, respectively. RPA and MS closures yield very similar results to that of the HNC and PY closures, respectively.} These mixed potentials are excellent at generating the target step-function \(g(r)\) pair structure up to densities very close to the singular limit, and, as expected, the low-\(k\) behaviors in \(S(k)\) indicate that the simulated compressibility drops rapidly with increasing \(\phi\).


\begin{figure}
  \includegraphics
  {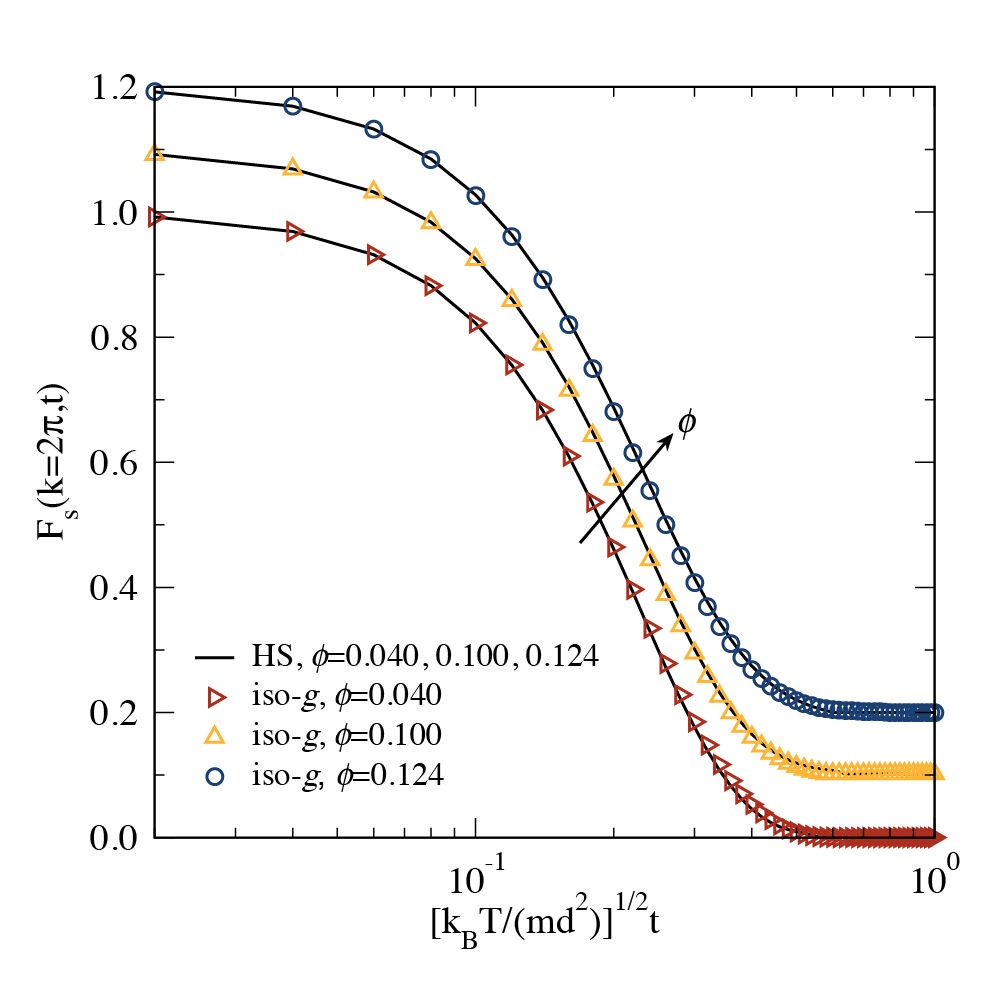}
  \caption{Self-intermediate scattering functions \(F_{\text{s}}(k,t)\) for selected packing fractions, where black lines correspond to HS systems and symbols correspond to systems with interactions \( u(r)\) designed to realize the iso-\(g\) process spanning \(0 \leq \phi \leq 1/8\).}
  \label{sch:Fskt-lowdens}
\end{figure}

Despite this precipitous drop in compressibility upon approach to $\phi_s$, we show in Fig.~\ref{sch:Fskt-lowdens} that the single-particle dynamics of this series of systems--as captured by the decay of the self-intermediate scattering function \(F_{\text{s}}\)--are virtually identical to those of HS systems at the same packing fractions. This result can be rationalized on the basis of thermodynamics and the form of the auxiliary interactions. First, we recall from Fig.~\ref{sch:auxiliary_range}c that for this range of \(\phi\), changes in both \(s_{2}\) and \(s\) upon suppressing otherwise emergent HS structuring are finite and quite small. Thus, the similarity in dynamics may have been anticipated based on the quasi-universal correlation between excess entropy and relaxation discussed above (though does not suggest whether either quantity is the \emph{better} predictor). Secondly, because the auxiliary interactions are smooth, monotonic, and increasingly long-ranged, they predominately establish a nearly additive renormalization of the system enthalpy relative to equidensity HS, introducing no new meaningful barriers to diffusion beyond those already present in the HS fluid. Accordingly, the rapidly decreasing compressibility near \(\phi_{\text{s}}\) does not coincide with any jamming-like transition, but simply with the growing mean-field-like repulsive tail \(e\) contribution to \(f\).

\subsection{Iso-\(g\) processes at higher packing fractions}
\label{results:all_isog}

\begin{figure}
  \includegraphics{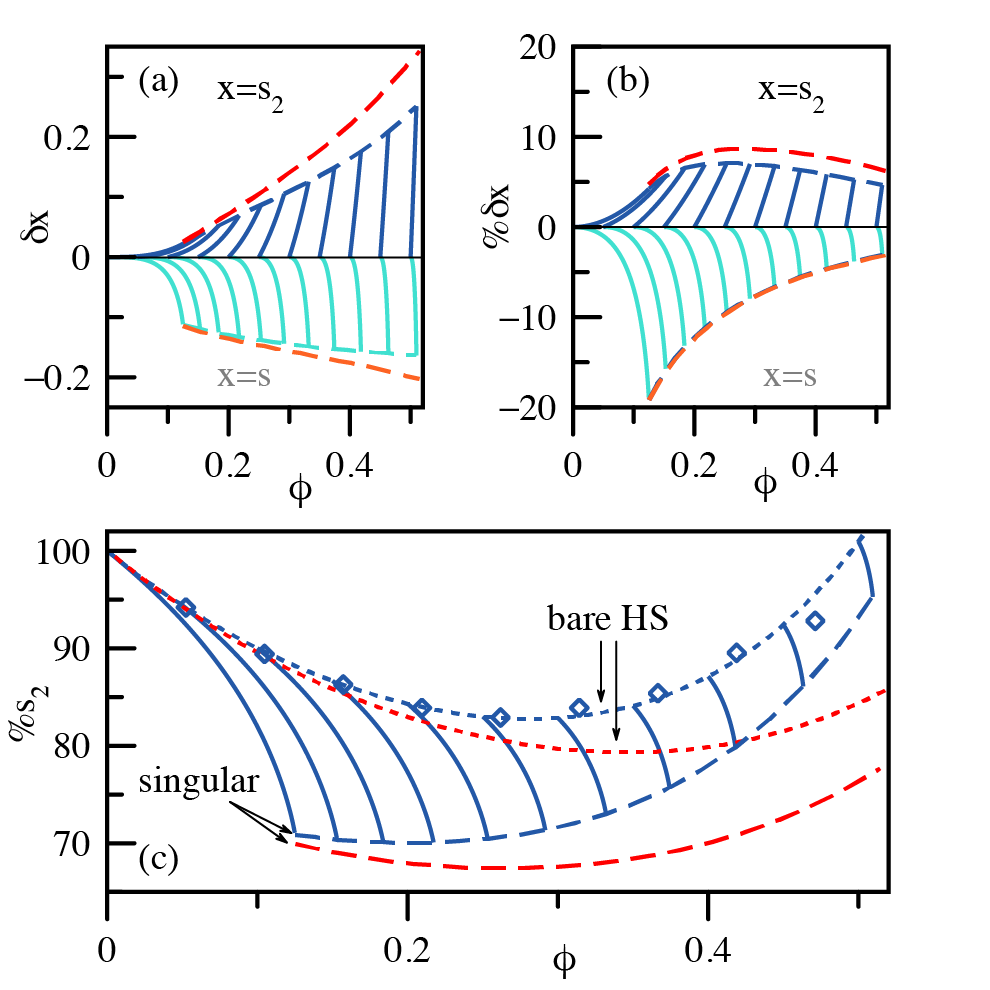}
  \caption{(a) \(\delta x \equiv x-x^{(\text{HS})}\) and (b) \(\%\delta x \equiv 100\times(x-x^{(\text{HS})})/|x^{(\text{HS})}|\) computed along a variety of hard sphere iso-\(g\) processes within the RPA (blue and teal) and HNC (red and orange) approximations. Solid curves correspond to the 11 individual iso-\(g\) processes (HNC not shown for clarity) while long dashed curves indicate the singular density values. (c) \(\%s_{2}\equiv 100 \times s_{2}/s\) for the 11 iso-\(g\) processes (solid) bare hard spheres (small dashed) and locus of singular density points (long dashed). Open diamond symbols are simulation results from ~\cite{s_s2_simulations}.}
  \label{sch:continuum_s2}
\end{figure}

As demonstrated for the step function iso-\(g\) process, both \(s_{2}\) and \(s\) undergo only minor departures from HS fluid values, despite divergence of the auxiliary potential range and some thermodynamic measures at \(\phi_{\text{s}}\). To test the generality of these observations, IET-based calculations for a variety of finite density iso-\(g\) processes are explored with \(0 \leq \phi_{0} \leq 0.50\) in 0.05 increments (11 iso-\(g\) processes). The first quantities of interest, shown in Fig.~\ref{sch:continuum_s2}a, are the two-body and full excess entropy differences between the iso-\(g\) processes and equidensity HS results, which we notate \(\delta x \equiv x-x^{(\text{HS})}\), where \(x=s_{2}\) or \(s\). While quantitative differences between the HNC and RPA theories emerge at higher density~\footnote[10]{This is presumably a result of the structural over-prediction of the underlying HS fluid within the HNC approximation.}, both theories predict all 11 iso-\(g\) processes furnish small overall changes in both \(s_{2}\) and \(s\). In addition to being small, the decrease in \(s\) at the various \(\phi_{\text{s}}\) points is nearly constant (roughly \(\delta s\approx-0.15\)), eventually being surpassed in magnitude by the change in \(s_{2}\) with increasing $\phi$. The latter observation implies higher densities require less of a decrease in total disorder to suppress otherwise emergent pair structure. To get a better sense for how large the thermodynamic departures are from HS along an iso-\(g\) process, we also consider \(\% \delta x \equiv 100\times(x-x^{(\text{HS})})/|x^{(\text{HS})}|\), which expresses the changes as percents relative to the absolute HS value. For \(s\), we find a monotonic decrease (in magnitude), which at the highest density iso-\(g\) process only achieves about a 4\% change. On the other hand, the \(s_{2}\) predictions are non-monotonic and even more weakly varying; evidently, the upper singular density permits only a meager 5-8\% change in \(s_{2}\) along a given HS iso-\(g\) process.

The quantities \(\delta x\) and \(\% \delta x\) lend insight into the change in \(s_{2}\) and \(s\) along an iso-\(g\) process, but it is also interesting to consider how an iso-\(g\) process \emph{repartitions} two- versus higher-body correlations relative to HS. To address this issue, we use the fact that \(s_{2}\) is the purely two-point contribution to \(s\) (i.e., \(s=s_{2}+s_{MB}\) where \(s_{MB}\) contains all ``many-body'' terms not trivially reducible to purely two-point correlations). Thus, the ratio (expressed as percentages) \(\%s_{2}=100 \times s_{2}/s\) intuitively expresses the percent of disorder due to two-body correlations. As shown in Fig.~\ref{sch:continuum_s2}c, \(\%s_{2}\) exhibits a non-monotonic variation that echoes that of the underlying behavior of the HS fluid (i.e., this is reflected by the singular points). For the lowest density iso-\(g\) process, the drop in \(\%s_{2}\) is fairly strong (100\% \(\rightarrow\) 70\% between \(\phi=\phi_{0} \rightarrow \phi=\phi_{\text{s}}\)), but higher density iso-\(g\) processes are much less dramatic. Overall, the HNC and RPA theories suggest that no HS-reference iso-\(g\) process is capable of swapping the dominant contributions towards ``disorder'' from two-body to many-body. Whether the same is true more generally for iso-\(g\) processes referenced against non-HS systems remains to be seen.

\begin{figure}
  \includegraphics{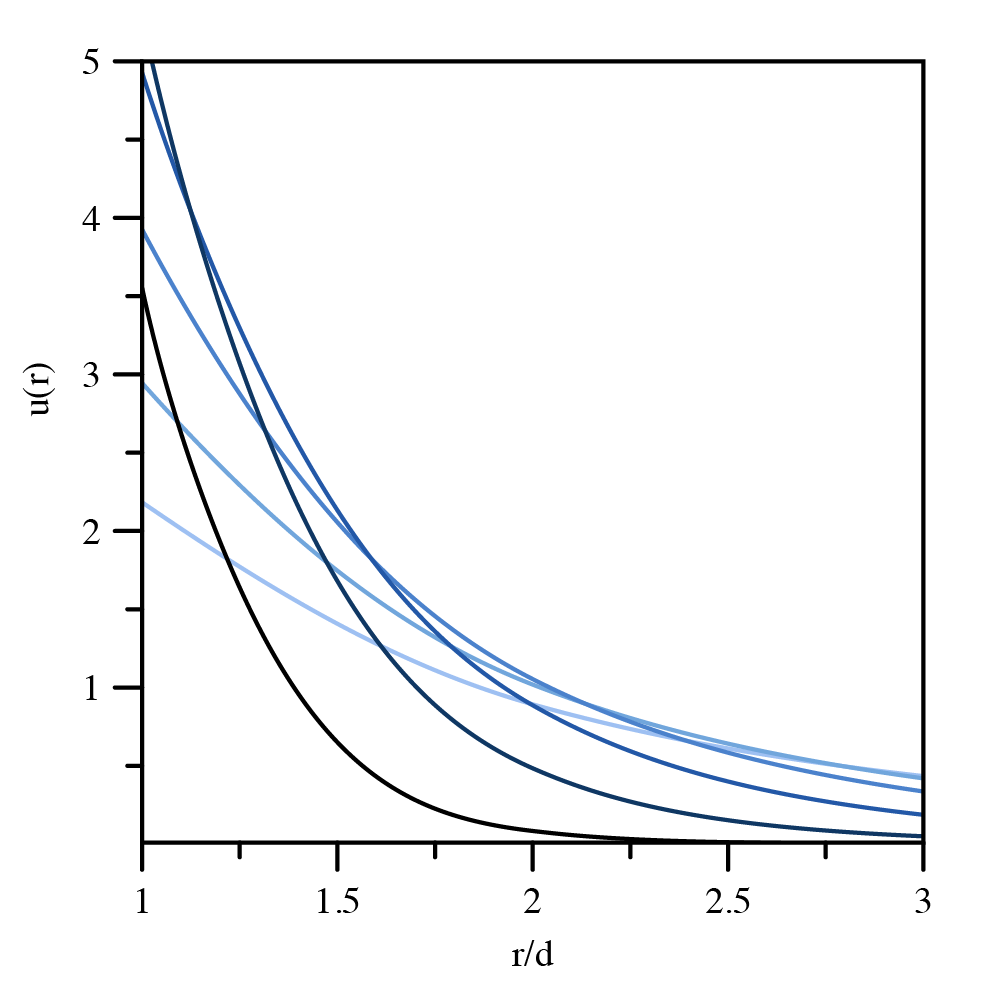}
  \caption{RPA predictions for the iso-\(g\) potentials at \(\phi=0.99\phi_{\text{s}}\) given the reference states \(\phi_{0}=0\), 0.10, 0.20, 0.30, 0.40, 0.50 (lightest to darkest). HNC predictions are nearly identical and are thus not shown for clarity.}
  \label{sch:continuum_potentials}
\end{figure}

In closing this section, we note that the auxiliary interactions required for the step-function iso-$g$ process are all ramp-like and purely-repulsive, a scenario that holds true for all of the HS iso-\(g\) processes. As shown in Fig.~\ref{sch:continuum_potentials}, the RPA theory predicts repulsive ramp-like potentials for each of the iso-\(g\) processes at \(\phi=0.99\phi_{\text{s}}\). The interaction range decreases with density, reflecting that higher density iso-\(g\) processes must approach closer (in density) to the singularity for symptoms of the divergence to be detected. All such potentials are slowly and smoothly varying, which is consistent with minimal changes in dynamics relative to the HS fluid upon approach to the respective singular packing fractions--in accord with the modest changes in the excess entropy metrics shown in Fig.~\ref{sch:continuum_s2}.

\subsection{Maximizing two-body disorder at arbitrary packing fraction}
\label{results:max_s2}

\begin{figure}
  \includegraphics{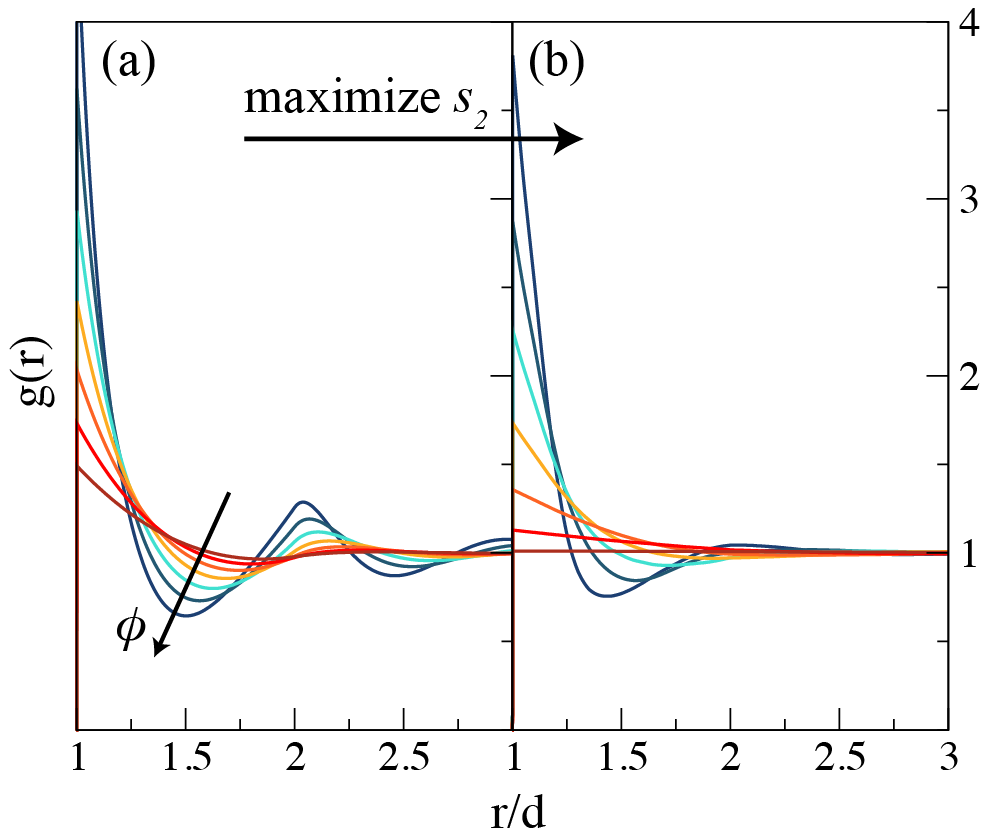}
  \caption{(a) Radial distribution functions of HS fluids for \(0.15 \leq \phi \leq 0.45\) in increments of \(0.05\). (b) Equi-density radial distribution functions optimized to maximize $s_{2}$.}
  \label{sch:optgr}
\end{figure}

Above \(\phi=1/8\), a unit step \(g(r)\) is not physically realizable because the corresponding \(S(k)\) is negative for some values of $k$. To find the \(g(r)\) of a fluid with hard cores that maximizes \(s_{2}\) subject to the constraint that \(S(k)\) be positive definite for all $k$, we use a genetic algorithm, beginning the optimization from the HS radial distribution functions (shown in Fig.~\ref{sch:optgr}a). The resulting equi-density radial distribution functions are shown in Fig.~\ref{sch:optgr}b. In general, the optimized structures bear some similarity to the HS fluid correlation functions, with muted features in the former, particularly at larger $r$. Just above the terminal density of the unit-step iso-$g$ process (at \(\phi = 0.15\)), the optimized \(g(r)\) is predominantly flat with a small, slowly varying ramp near contact. As $\phi$ increases, more structuring appears: the value of \(g(r)\) at contact grows, and a depletion region following the first coordination shell becomes apparent that grows in magnitude and shifts to lower $r$ with increasing $\phi$. At the highest values of $\phi$, a second coordination shell appears. All $s_2$-maximized pair-correlation functions are relatively short-ranged (much more so than the HS counterparts).

\begin{figure}
  \includegraphics{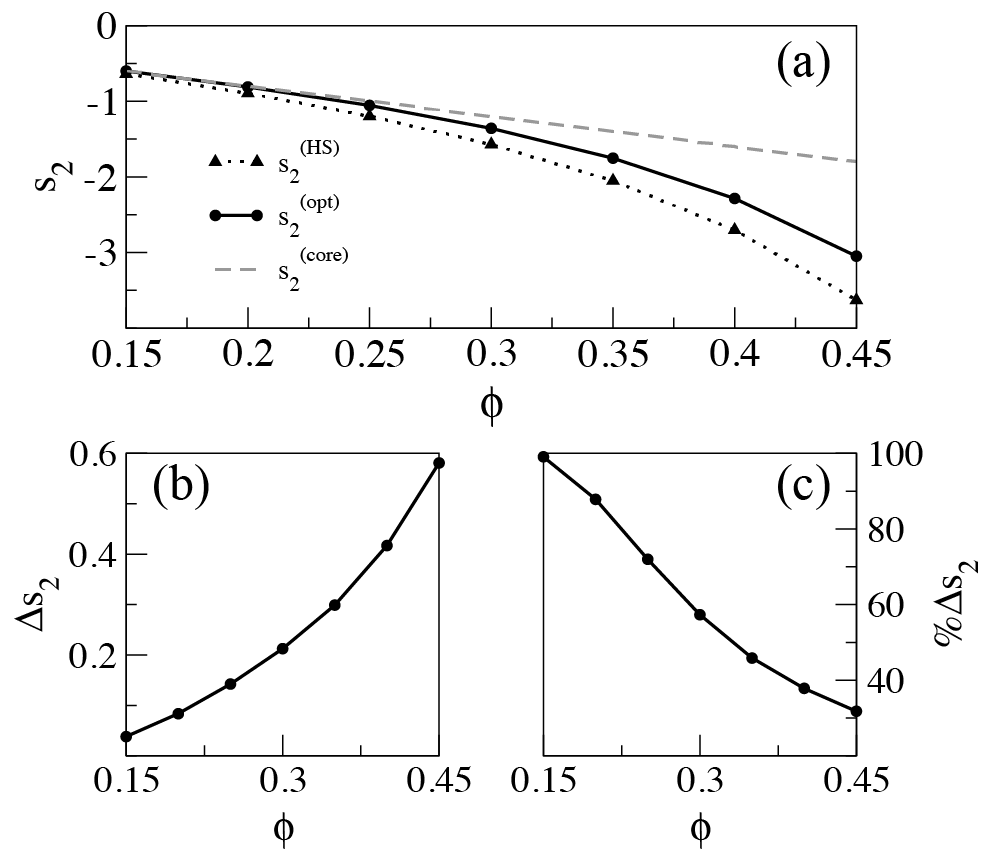}
  \caption{(a) Two-body excess entropy for the HS fluid ($s_{2}^{(\text{HS})}$), maximal $s_{2}$ subject to the structure factor realizability constraint ($s_{2}^{(\text{opt})}$), and $s_{2}$ of an unrealizable unit-step radial distribution function ($s_{2}^{(\text{core})}$) as a function of packing fraction $\phi$. (b) Increase in $s_{2}$ relative to hard spheres: $\Delta s_{2} = s_{2}^{(\text{opt})}-s_{2}^{(\text{HS})}$ (c) Percentage of $s_2$ gap between $s_{2}^{(\text{core})}$ and $s_{2}^{(\text{HS})}$ recovered by the fluid upon $s_2$ optimization.}  
  \label{sch:opts2}
\end{figure}

Fig.~\ref{sch:opts2}a shows \(s_{2}\) for the HS fluid, $s_2$ from the optimized radial distribution functions (with the $S(k)$ positivity constraint), and $s_2$ of an (unrealizable) fluid with a unit-step radial distribution function. Of course, at lower densities where the \(g(r)\) most closely resembles a unit step, the bulk of \(s_{2}\) comes from the core. But non-hard core contributions due to the coordination shells grow in with increasing $\phi$. Together, \(s_{2}\) for the HS radial distribution function and \(s_{2}\) for the unit-step radial distribution function provide a range in which the optimized \(s_{2}\) must lie. In absolute terms, at the higher densities, the optimization has a greater effect to maximize \(s_{2}\). We show this in Fig.~\ref{sch:opts2}b, where the increase in \(s_{2}\) from the HS to the optimized radial distribution function is plotted as a function of $\phi$. However, if we consider the change in \(s_{2}\) upon optimization as a percentage of what could have otherwise been achieved without the $S(k)>0$ realizability constraint, we see that this quantity decreases with $\phi$; see Fig.~\ref{sch:opts2}c. In other words, it is primarily the realizability constraint that limits pair disorder at higher $\phi$.

\begin{figure}
  \includegraphics{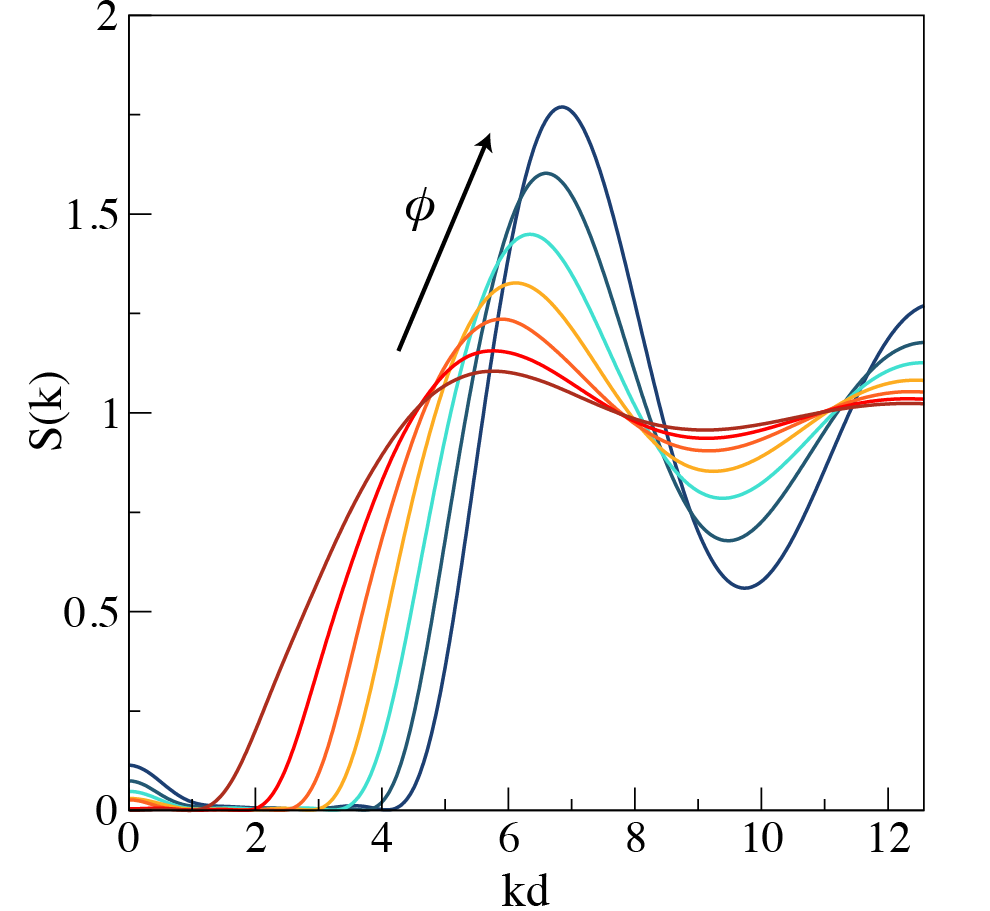}
  \caption{Structure factors corresponding to the optimized \(g(r)\) profiles in Fig. 6. Note that oscillations at low \(k\) are minor artifacts attributable to the numerical optimization process, which can in principle be refined via further iterative optimizations.}
  \label{sch:optSk}
\end{figure}

The structure factors corresponding to the optimized radial distribution functions, shown in Fig.~\ref{sch:optSk}, are characteristic of disordered, stealthy, hyperuniform materials as they are transparent (do not scatter) over a finite range at low \(k\)~\cite{stealthy_hyperuniform_I,stealthy_hyperuniform_II}. Due to the constrained optimization, any solution above the terminal density must have at least one value of $k$ where $S(k)$ is zero; we find a whole range of low $k$ values, beginning at \(k = 0\), where \(S(k) \approx 0\). Small oscillations above zero in the low $k$ region are due to the inexactness of the numerical optimizations and decrease in magnitude as the optimization progresses (not shown). The range of $k$ values for which \(S(k) \approx 0\) increases with $\phi$. This suggests an intimate relation between maximizing $s_{2}$ and the fabrication of hyperuniform, ``stealthy'' materials that possess minimal scattering at low wavevectors. This is a potentially useful relationship because optimizing materials in real space (via interactions and radial distribution functions) allows for a simpler encoding of the hard-core constraint (or other real space constraints) than does directly manipulating $S(k)$. These calculations do not strictly prove the existence of such stealthy hard-core packings but, if realizable by explicit construction, would provide a complement to the library of currently existent \emph{point} particle stealthy packings~\cite{stealthy_hyperuniform_I,stealthy_hyperuniform_II}.   

%
\section{Conclusions}
\label{conclusions}

Hard spheres are the archetypical model for a diverse array of systems at both the macroscale and microscale, but universal to all is the role of the hard core to restrict the available phase space and to limit the amount of disorder that is realizable. As discussed in this work, thermalized hard spheres are maximally disordered when the excess entropy, \(s\), is considered; introduction of any other interaction beyond the hard core necessarily lowers \(s\). However, maximizing the disorder stored in the two body correlations, \(g(r)\), is non-trivial. As our metric for pair disorder, we employed the two-body excess entropy, \(s_{2}\), which naturally arises from the multi-body expansion of \(s\). 

To investigate the role of increasing \(s_{2}\), we explored a variety of HS iso-\(g\) processes: from a given starting density, auxiliary interactions are introduced as the density is increased so that the initial HS \(g(r)\) is maintained. Such processes terminate when the singular density is reached; these singular points are characterized by incompressibility and hyperuniformity--features commonly associated with jamming or close-packed states. Unlike jammed packings though--which would exhibit negative divergences in \(s\) and \(s_{2}\) at the jamming transition--our IET calculations predict the iso-\(g\) singularities have \(s\) and \(s_{2}\) values close to equidensity hard spheres. This suggests the singular points are of virtually the same mobility as underlying hard spheres--a prediction we confirmed with molecular dynamics simulations for the step function \(g(r)\) (or zero starting density) iso-\(g\) process. Incompressibility at the singular densities is solely a manifestation of the divergent-ranged interactions required to maintain the lower-density radial distibution function, as opposed to a literal jamming transition. The smooth, slowly varying nature of the interactions seems insufficient for generating the frustrated energy landscape needed for slow glassy dynamics.

As a consequence of increasing the pair disorder along each iso-\(g\) process, stronger correlations are imbued into the \(n>2\) body structure. IET predicts that the increase in \(s_{2}\) is comparable is magnitude to the decrease in \(s\), meaning that the increase in \(s_{2}\) is accompanied by roughly twice as large of a decrease in the many body correlations in \(s=s_{2}+s_{MB}\). Despite transferring order from $s_{2}$ to $s_{MB}$, \(\gtrsim 70\%\) of \(s\) of the iso-$g$ process is predicted to result from \(s_{2}\). This effect is rather mild, particularly considering that bare hard spheres have a contribution to $s$ from \(s_{2}\) of \(\gtrsim 80\%\). Thus, we conclude that HS iso-\(g\) processes are not capable of swapping the dominant contribution to $s$ from two- to many-body. Whether other systems can be found where \(s_{MB}>s_{2}\) is an interesting avenue for future research.

While iso-\(g\) processes produce hard-core fluids of augmented \(s_{2}\) relative to pure hard spheres, a literal maximization of \(s_{2}\) subject to the constraint of a positive-definite \(S(k)\) was performed to uncover the true limits of pair disorder. Below \(\phi = 1/8\), the solution is the trivial step function \(g(r)\) (degenerate with iso-\(g\) processes beginning at \(\phi_0=0\)). Above \(\phi = 1/8\), some degree of structure in \(g(r)\) must remain for realizability. As density is increased, the optimized $s_{2}$ is increasingly greater than the $s_{2}$ of equidensity hard spheres; however, this is a reflection of the greater available (i.e, non-hard core) $s_{2}$ at higher densities for hard spheres. Expressed as a percentage of $s_{2}$ available to the optimization, we see that relative change in $s_{2}$ actually decreases as a function of density, reflecting that physical realizability is a more difficult constraint to satisfy at higher densities. Finally, we found that the resulting radial distribution functions are hyperuniform, with a range in low $k$ corresponding to $S(k) \approx 0$ that broadens with increasing density, indicating a relationship between maximizing pair disorder and so-called ``stealthly'' materials. This relationship provides a convenient route to directly generate stealthy packing pair structure in a system of particles with hard cores. Explicit construction of such packings is an interesting avenue for future research.

\section{Acknowledgements}
This work was partially supported by the National Science Foundation (1247945) and the Welch Foundation (F-1696). We acknowledge the Texas Advanced Computing Center (TACC) at The University of Texas at Austin for providing HPC resources

\section*{Appendix A. Asymptotic behavior of potentials}
\label{gen_isog_math}

For completeness, this section provides a concise review of general analytical results regarding any iso-\(g\) process, though the presented mathematical formalism follows a slightly different path than in the current literature~\cite{isog_and_hyperuniform}. 

General analytical results regarding any iso-\(g\) process rely on approximating \(c(r)\) and exploiting the intimate relation it possesses with \(u(r)\), in particular, the physical interpretation of \(k_{B}Tc(r)\) as a renormalized pair potential between particles and the exact result, \(\lim_{r\to\infty} c(r)=- u(r)\). From this relation it is clear that knowledge of the asymptotic \(c(r)\) behavior--extractable from the OZ relation alone--provides the asymptotic form of \(u(r)\). All three of the considered closures (Equations ~\ref{eqn:HNC_closure}-\ref{eqn:RPA_closure}) obey this limit via the trivial relation, \(\lim_{r\to\infty} g(r)=1\)--an expected consistency. 

Analytical statements regarding the long range \(c(r)\) and \(u(r)\) behavior are most readily found near the singular density, \(\rho_{\text{s}}\), at which \(S(k)\equiv 1+\rho_{\text{s}} \hat{h}(k)=0\) for some \(k=k_{\text{s}}\). At \(k_{\text{s}}\), a divergence in \(\hat{c}(k)\) is produced since
\[1-\rho \hat{c}(k)=\dfrac{1}{1+\rho \hat{h}(k)}\tag{A1}\label{eqn:ck_divergence}\]
yielding \(c(k_{\text{s}})\propto (\rho_{\text{s}}-\rho)^{-1}\) since \(\rho_{\text{s}}=-1/\hat{h}(k_{\text{s}})\). From Equation~\ref{eqn:ck_divergence}, analytical results regarding the nature of the divergence, both in Fourier and real space, are elucidated through a Taylor expansion of \(\hat{h}(k)\) in the denominator about \(k_{\text{s}}\) yielding
\[1-\rho \hat{c}(k)\approx \dfrac{1-\rho \hat{c}(k_{\text{s}})}{1+\xi_{k_{\text{s}}}^{2}(k-k_{\text{s}})^{2}}, \ \ k\approx k_{\text{s}}\tag{A2}\label{eqn:ck_lorentzian}\]
and
\[\xi_{k_{\text{s}}}^{2}\equiv \dfrac{\rho \hat{h}^{\prime \prime}(k_{\text{s}})}{2[1+\rho \hat{h}(k_{\text{s}})]}\tag{A3}\label{eqn:correlation_length_I}\]
where the primes indicate the order of differentiation, and  \(\xi_{k_{\text{s}}}\) is the auxiliary potential range. Interestingly, an alternative expression (though identical at the singular density) can be derived for \(\xi_{k_{\text{s}}}^{2}\) by working with \(\hat{c}(k)\) directly instead of \(1-\rho \hat{c}(k)\) 
yielding
\[\xi_{k_{\text{s}}}^{2}\equiv \dfrac{- \hat{h}^{\prime \prime}(k_{\text{s}})}{2\hat{h}(k_{\text{s}})[1+\rho \hat{h}(k_{\text{s}})]}\tag{A4}\label{eqn:correlation_length_II}\]
Both results furnish a divergence, \(\xi_{k_{\text{s}}}\propto(\rho_{\text{s}}-\rho)^{-1/2}\), and together they appear to serve as bounds on the exact \(\xi_{k_{\text{s}}}\) (discussed in Section~\ref{results:step_isog}). Greater physical interpretation for \(\xi_{k_{\text{s}}}\) as the auxiliary interaction range comes from the Fourier transformed versions of Equation~\ref{eqn:ck_lorentzian}, which for \(k_{\text{s}}=0\) yields
\[\lim_{r\to\infty}c(r)=- u(r)=\dfrac{1-\rho \hat{c}(0)}{\rho 4\pi \xi_{k_{\text{s}}}^{2}}\dfrac{\text{exp}[-r/\xi_{k_{\text{s}}}]}{r}\tag{A5}\label{eqn:yukawa_k0}\]
Arbitrarily close to the singular density
\[\lim_{\rho \to \rho_{\text{s}}}\lim_{r\to\infty}c(r)=- u(r)=\dfrac{\hat{c}(0)}{4\pi \xi_{k_{\text{s}}}^{2}r}\tag{A6}\label{eqn:power_law_k0}\]
which is well defined as \(\lim_{\rho \to \rho_{\text{s}}} \hat{c}(0)/\xi_{k_{\text{s}}}\) remains finite.



\begin{acknowledgments}
This work was partially supported by the National Science Foundation (1247945) and the Welch Foundation (F-1696). We acknowledge the Texas Advanced Computing Center (TACC) at The University of Texas at Austin for providing HPC resources.
\end{acknowledgments}



%


\end{document}